\documentclass[a4paper,11pt]{article}
\usepackage{pos}
\usepackage{subcaption}
\usepackage{rviewport}
\usepackage{wrapfig}
\usepackage{enumitem}
\usepackage{multicol}
\usepackage{setspace}

\title{{Detection of Spaceborne Lasers\\ with the Pierre Auger Observatory}}
\ShortTitle{Detection of Spaceborne Lasers with the Pierre Auger Observatory}

\author*[a]{Michael Unger}
\affiliation[a]{Institute for Astroparticle Physics, Karlsruhe Institute of Technology (KIT), 76131 Karlsruhe, Germany}

\onbehalf{for the Pierre Auger Collaboration$^b$}
\affiliation[b]{Observatorio Pierre Auger, Av.\ San Mart{\'\i}n Norte 304, 5613 Malarg\"ue,
Argentina\\ Full author list: {\rm\url{https://www.auger.org/archive/authors_icrc_2025.html}}}

\author[c]{and Oliver Lux}
\author[c]{Oliver Reitebuch}
\affiliation[c]{Deutsches Zentrum für Luft- und Raumfahrt (DLR German Aerospace Center), Institute of Atmospheric Physics, 82234 Oberpfaffenhofen, Germany}

\emailAdd{spokespersons@auger.org}

\abstract{The detection of side-scattered ultraviolet light from spaceborne lasers with fluorescence telescopes of cosmic ray observatories offers unique opportunities for systematic studies of the aerosol content of the local atmosphere. It also enables the validation of the optical calibration of the telescopes. Additionally, these observations provide valuable ground-based monitoring of the performance of the scientific instruments aboard satellites used for Earth climate observation.

Here, we report on results from the reconstruction of laser shots from the spaceborne lidar instrument ALADIN aboard the Aeolus satellite in 2019, 2020 and 2021. Furthermore, we present initial observations of laser shots from ATLID, the atmospheric lidar of the EarthCARE satellite, launched in 2024. EarthCARE’s orbit is particularly well-suited for enabling laser detection within a few days at both the Pierre Auger Observatory and the Telescope Array Experiment, facilitating a relative calibration of the energy scales of these observatories.%
}

\ConferenceLogo{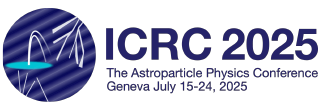}

\FullConference{39th International Cosmic Ray Conference (ICRC2025)\\
 15–24 July 2025\\
Geneva, Switzerland\\}

\begin{document}
\maketitle

\section{Introduction}

Cosmic-ray observatories such as the Pierre Auger Observatory and the
Telescope Array  Experiment detect air showers with large-aperture ultraviolet
(UV) fluorescence telescopes. To convert raw light tracks into
accurate shower profiles, the state of the atmosphere needs to be
known precisely, especially the density of aerosols that scatter and
absorb the UV light. The standard tool to perform these measurements is
to detect the side-scattered light from ground-based
lasers~\cite{HiRes:2005omc, Fick:2006faa,   PierreAuger:2013dlv,TelescopeArray:2011mma} and
ground-based lidar systems~\cite{Tomida:2011cb, PierreAuger:2012rhm, Rizi:2019hba}.

Space-based lasers have been observed previously by
VERITAS~\cite{VERITAS:2023ouz} and
TAIGA-HiSCORE~\cite{Porelli:2017blz,Porelli:2021jvi}.  Recently,
Earth-observation satellites have begun to carry high-energy UV lidars
to profile winds, clouds, and aerosols on a global scale. The first
mission equipped with an UV laser was Aeolus in 2018, followed up by
EarthCARE in 2024. The lidar beams from these satellites are bright enough to
be detected at astroparticle observatories that were built for
detecting the far dimmer UV fluorescence and Cherenkov light from air
showers.  Here we report on the observations of UV laser tracks from
Aeolus and EarthCARE with the Pierre Auger
Observatory~\cite{PierreAuger:2021obf}.

In these proceedings, we discuss how the detection of space-based
lasers from the ground can be used to calibrate both space- and
ground-based instruments. First, we summarize how the fluorescence
telescopes of the Pierre Auger Observatory were used to ground-truth
the lidar of the Aeolus satellite. We then present early results from
the observation of laser shots from EarthCARE. The orbit of this
satellite allows for a year-round observation, making it particularly
useful to cross-check the standard aerosol
reconstruction. Furthermore, the laser tracks of EarthCARE can be
observed at the Pierre Auger Observatory in Argentina and the
Telescope Array Experiment in USA within a few days during the same
moon cycle. This opens the possibility for a direct cross-calibration
of the energy scale of the two cosmic-ray observatories.

Together, these studies show that the collaboration between cosmic-ray
observatories and space-based climate-sensing missions is mutually
beneficial for understanding and calibrating both instruments.

\begin{figure}[htbp]
  \begin{minipage}[t]{0.48\textwidth}
\centering
\includegraphics[width=\linewidth]{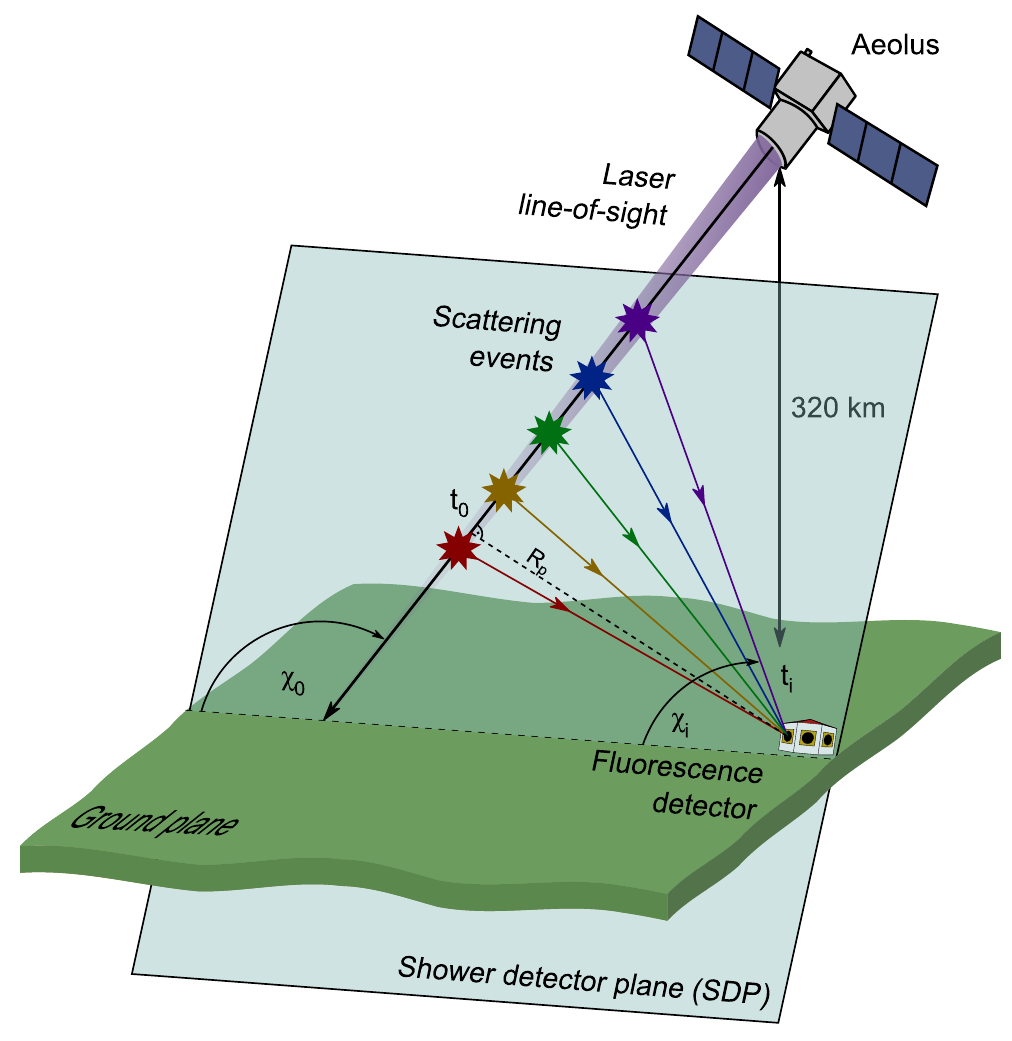}
\caption{Geometry of the Aeolus laser beam being detected by one of
  the four fluorescence detectors of the Pierre Auger Observatory~\cite{2024Optic..11..263T}.}
\label{fig:aeolus_pao_geometry}
  \end{minipage}%
  \hfill
  \begin{minipage}[t]{0.48\textwidth}
\centering
\includegraphics[width=0.95\linewidth]{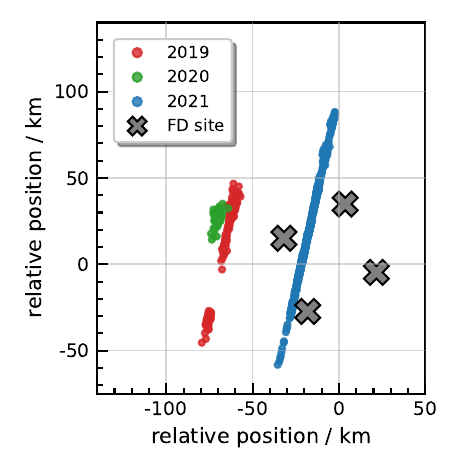}
\caption{Measured laser impact point at an altitude of 1400 m for
  three sample Aeolus overpasses in the years 2019, 2020, and 2021~\cite{2024Optic..11..263T}.}
\label{fig:combined_tracks}
  \end{minipage}
\end{figure}

\section{Observation of Aeolus}
The ESA Aeolus satellite (2018-2023) carried the UV Doppler wind lidar
ALADIN. After the launch, the atmospheric back-scatter signal of the
instrument declined and the identification of the root cause, losses on the emit or receive
path, could not be unambiguously clarified. In mid-2019, the
ultraviolet pulses were serendipitously recorded by the telescopes of
fluorescence detector (FD) of the Pierre Auger Observatory. The 27
Schmidt-optics telescopes are optimised to detect the 300-400 nm
fluorescence light from cosmic-ray air showers.

Since the FD is capable of detecting pulses of faint UV light, a
measurement of the Aeolus laser using the telescopes of the Pierre
Auger Observatory is possible, as illustrated in
Fig.~\ref{fig:aeolus_pao_geometry}. Due to the nature of the
sun-synchronous dusk-dawn orbit of Aeolus, it always passed close to
the sunrise or sunset over any given point on the surface of the
Earth. This limited the opportunities for measurements with the
FD. Additionally to the aforementioned limitations by the moon-cycle,
a measurement could only take place if the satellite passage time fell
within the astronomical night, which occurred only during the
southern-hemisphere winter months, i.e.\ between May and
August. Furthermore, the orbit had a 7-day repeat cycle. Therefore, a
measurement could only be performed once per week. Overall this resulted in
up to six observations of the Aeolus laser per year, when clouds were
not preventing the visibility.

These detections initiated a ground-based monitoring of the space laser
and provided independent ``ground truth'' for both impact point (ground
track) and pulse energy. In Ref.~\cite{2024Optic..11..263T} we
reported on three high-quality overpasses (3 Aug 2019, 27 Jun 2020, 17
Jul 2021) selected from 16 detections between 2019-2021, all taken in
moonless, clear-sky winter nights when Aeolus crossed the array during
astronomical darkness.

\subsection{Geometrical reconstruction of the laser ground track}

\begin{figure}[htbp]
  \centering
  \begin{subfigure}[t]{0.4\textwidth}
\includegraphics[width=\linewidth]{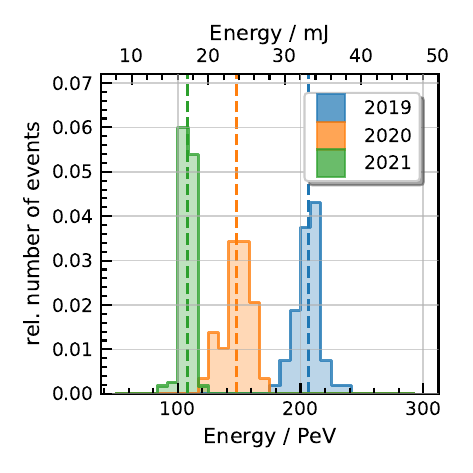}
\caption{Reconstructed energies for three sample Aeolus overpasses in
  2019, 2020, and 2021. The average energy per overpass is marked by
  the dashed line.}
\label{fig:combined_ehists}
  \end{subfigure}
  \hfill
  \begin{subfigure}[t]{0.57\textwidth}
\includegraphics[width=\linewidth]{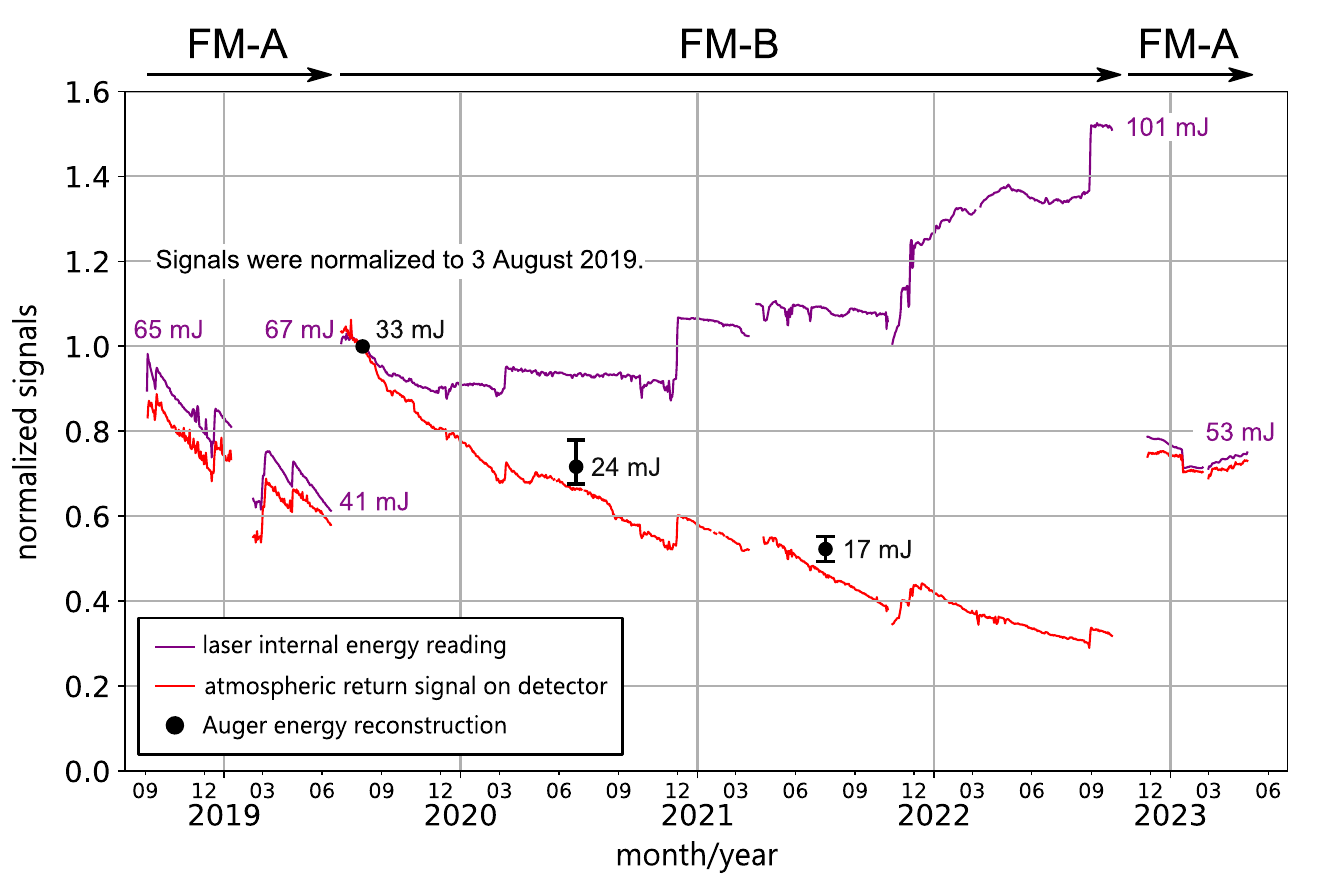}
\caption{Signal evolution of the Aeolus instrument: The purple curve
  represents the laser energy as measured at the output of the
  respective transmitter. The red curve denotes the atmospheric return
  signal that is detected on the Rayleigh receiver under clear-air
  conditions. The black dots indicate the Auger measurements, error
  bars denote statistical uncertainties.  }
\label{fig:aeolus_signal_evolution}
  \end{subfigure}
  \caption{Reconstruction of the Aeolus laser energy~\cite{2024Optic..11..263T}.}
  \label{fig:combined}
\end{figure}

The Aeolus laser was visible in the FD due to the scattering of the
laser beams with air molecules in the atmosphere. Similarly to the
fluorescence light of showers, this light could be measured by the
telescopes, creating an image of the laser beam in the camera
constituting a line of pixels.  This served as the basis for the
geometrical reconstruction of the laser beam. If the beam was seen by
only one telescope, the geometry could be obtained only by a so-called
monocular reconstruction. Here we use an improved,
zenith-angle-constrained, monocular reconstruction to determine the
geometry of the laser track~\cite{Knapp:21}. We first determined the
average of the arrival direction of the laser beam from the combined
data set, and then fix the direction during reconstruction to this
average value, because the arrival direction of the laser beam does
not change between the orbits. This reduces the number of geometry
parameters from five to four, and it removes the usual
near-degeneracies for the geometry within the shower-detector-plane
(see \ref{fig:aeolus_pao_geometry}).

The reconstructed impact points
of the laser at an altitude of 1400\,m (WGS84) for three sample
overpasses in the years 2019, 2020, and 2021 is shown in Fig.~\ref{fig:combined_tracks}. Notable is the change
of the laser ground position for the year 2021 further to the East
after an adjustment of the satellite orbit, which was obtained via a
dedicated maneuver to allow for better measurements of the laser beam
by the Observatory.  The passages during the former two years happened
at larger distances and thus fewer events and shorter tracks were
observed.

When comparing the Aeolus ground track as determined by the Pierre
Auger Observatory with the positions reported in the Aeolus data
products, a horizontal offset of approximately 0.075$^\circ$ in
longitude (6.8~km) became evident. This discrepancy was traced to an
error in the Aeolus Level 1A (L1A) processor responsible for
calculating the geolocation of Aeolus observations. In particular, the
geolocation routines require time information along with a corresponding
identifier specifying whether the time is in UTC (Coordinated
Universal Time), GPS (Global Positioning System), or TAI
(International Atomic Time). In two instances, an incorrect
combination of time and identifier was passed to the geolocation routines,
leading to slightly erroneous geolocation calculations.

The Auger observations of the Aeolus ground track also enabled an
evaluation of the pointing accuracy and precision of the
satellite. Auger data are available for each individual Aeolus laser
pulse, whereas the Aeolus ground track in the L1A product is reported
only at the measurement level, which corresponds to an average over 30
pulses. To enable a direct comparison, individual pulse times were
reconstructed from the laser pulse frequency of 50.5~Hz and the
measurement centroid times reported in the L1A product. The
corresponding ground-track positions were then derived by time-based
interpolation of the longitude and latitude values provided in the L1A
data.

A pointing accuracy of 0.06~km along track and 0.82~km across track
was determined. The resultion, expressed in 2$\sigma$, is 1.28~km
along track and 0.93~km across track. These values represent a
combination of Aeolus pointing errors, interpolation uncertainties,
and Auger measurement errors, and therefore constitute upper limits on
the true Aeolus pointing accuracy and precision. Nonetheless, they lie
well within the Aeolus mission requirement of 2.0~km (2$\sigma$) for
horizontal geolocation, defined at the observation level averaged over 600 pulses.

\subsection{Pulse-energy reconstruction}

The fluorescence telescopes are absolutely calibrated with a 2.5~m
diameter drum light source~\cite{Brack:2013bta}. The uncertainty of
the absolute calibration of the telescopes, including also
contributions from the light-collection efficiency, reconstruction
bias, molecular atmosphere, multiple scattering and the long-term
calibration stability is 13~\%~\cite{Dawson:2020bkp,Verzi:13}.

For every laser event the number of photo-electrons in the camera is converted to the received photon
flux, corrected for atmospheric transmission (Rayleigh and aerosol optical depth) and
distance from the telescope to the track. Taking furthermore into account the attenuation along the
laser track, pulse energy at the spacecraft exit aperture is obtained. Simulations indicate a small
($\leq 3.7\%$) negative reconstruction bias, which is removed in the final numbers.

An energy reconstruction was performed for each individual laser
track. Using the same three representative nights as for the geometry
reconstruction, spanning the years 2019, 2020, and 2021, the resulting
energy distributions are shown in Fig.~\ref{fig:combined_ehists}. For
improved readability, the histograms were normalized to display
relative rather than absolute event counts per bin. As can be seen, a
steady decrease in reconstructed energy is observed over the years.

Figure~\ref{fig:aeolus_signal_evolution} presents the temporal
evolution of the ALADIN signal over the course of the Aeolus
mission. The laser energy measured with a laser-internal photodiode is
shown in purple, while the atmospheric return signal recorded by the
onboard receiver is plotted in red. The latter is derived from the
Rayleigh channel and restricted to observations under clear-air
conditions to ensure predominantly molecular backscatter. Energy
estimates obtained from the Auger laser-beam reconstruction are shown
as black dots. To enable a direct comparison of signal evolution at
different stages along the laser path (i.e., at the transmitter
output, after atmospheric propagation as seen by Auger, and at the
receiver) the signal levels are normalized to their respective values
on 3 August 2019, the date of the first fiducial Auger measurement. In
good approximation this normalization cancels out the systematic
calibration uncertainty of the Auger energy estimates (13\%), leaving
only the statistical uncertainty. The uncertainty of the first
measurement (normalized to unity) propagates into the relative
uncertainties of the subsequent data points. The periods where
different lasers were used, laser FM-A (Flight Model A) and FM-B, are
indicated as arrows at the top of the plot.

The evolution of the energies reconstructed by the Auger Observatory
closely follows the declining signal levels measured by ALADIN's
receiver between 2019 and 2021. This agreement suggests that the
observed signal loss during the FM-B laser operation originated along
the emit path, i.e., between the laser output and the telescope. The
Auger observations from July and August 2021 played a key role in
supporting the root-cause analysis of this degradation and ultimately
informed the decision to switch back to the FM-A laser in November
2022. Despite its lower output energy at the time of switchover
(53~mJ instead of 101~mJ), the transition to FM-A resulted in a 2.2-fold increase in
atmospheric return signal, bringing it back to the level seen during
the initial FM-A period in early 2019. This full recovery of signal
strength confirmed that the losses had occurred in the optical
components unique to FM-B, most likely within the relay optics, which
guide the FM-B beam onto the nominal optical axis. The specific loss
mechanism remains under investigation. Current studies focus on
laser-induced contamination, laser-induced damage, and bulk darkening
of the optical elements as the most likely causes.

\begin{figure}[t]
  \centering
\includegraphics[width=0.9\linewidth]{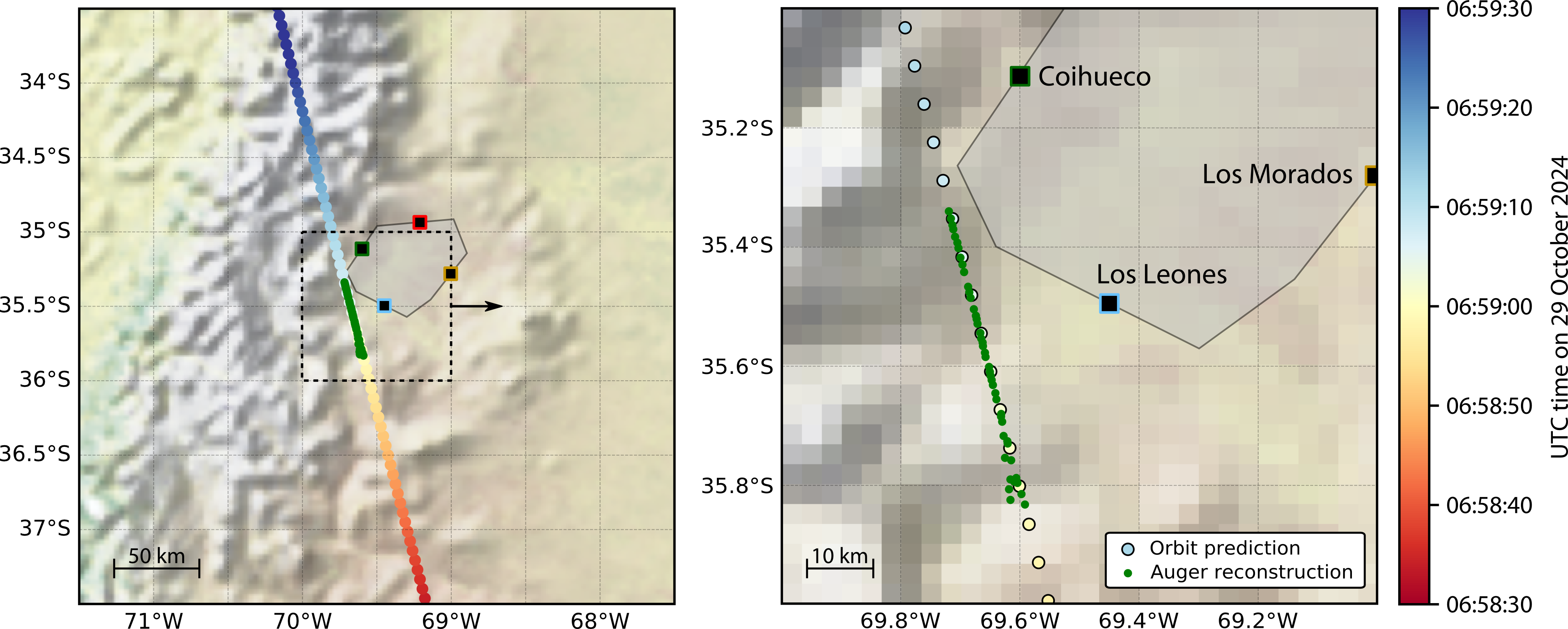}
  \caption{Reconstructed and predicted impact points of the EarthCARE
    laser at an altitude of 1400~m for the western pass near the Pierre Auger Observatory on 29 October 2024.}
  \label{fig:earthcare}
\end{figure}

Interestingly, the energy reductions observed by the Auger Observatory
in 2020 ($-28 \%$) and 2021 ($-48 \%$) relative to 2019 are slightly
smaller than the corresponding losses measured at the ALADIN detectors
($\-34\%$ and $-53\%$). This discrepancy suggests an additional
degradation mechanism in the receive path, most likely a clipping of
the atmospheric return signal at the field stop of the receiver.

It is also noteworthy that the absolute laser energy in 2019
reconstructed by the Auger Observatory ($\approx 33$~mJ) is lower than
the value of 48~mJ expected at the telescope output. This indicates
additional unaccounted losses in the emit path. An even larger discrepancy
of about a factor of two compared to pre-launch simulations was
already observed after launch in the atmospheric backscatter signals
from the Rayleigh channel~\cite{Reitebuch_ILRC}.

\section{Observation of EarthCARE}
The Atmospheric Lidar (ATLID) on ESA’s EarthCARE mission began routine
UV (355 nm) operations in mid-2024. Within weeks, its $\sim$35~mJ, 51~Hz
pulses were detected by the fluorescence telescopes of the Pierre
Auger Observatory and, shortly thereafter, by the fluorescence
detector of the Telescope Array  Experiment~\cite{tfCom} and the Cherenkov
telescopes of VERITAS~\cite{gmCom}.

Unlike Aeolus, whose dawn-dusk orbit limited visibility to a handful
of winter nights, EarthCARE offers two nighttime passes over each site
per 25-day repeat cycle -- one ``western'' and one ``eastern''. For
Auger, the western and eastern overpasses together total 28 per year;
roughly half of these fall into moonless measurement periods, so about
14~EarthCARE observations are feasible annually.  Local overpass time
is around 04:00, meaning that even at mid-summer the laser is visible
before astronomical dawn.
For the Telescope Array  Experiment, the western pass is the closer of the two,
occurring at about 03:00 local time, five days after the western Auger
overpass and four days before the eastern one.

Using the same laser-reconstruction software developed for Aeolus, we
reconstruct the ground track of the laser once again. Here we use an
angle of $3^\circ$ off nadir for the zenith-angle-constrained,
monocular reconstruction.
As can be seen in Fig.~\ref{fig:earthcare}, the reconstructed impact points for the overpass on the 29th of October 2024 are in good agreement with the ones expected from the EarthCARE orbit.

Beyond providing a well-timed global light source for fluorescence
telescopes, EarthCARE data can improve and validate the standard Auger
aerosol determination.  By deriving vertical aerosol optical depth
(VAOD) from EarthCARE laser tracks one can cross-check the VAOD
profile reconstruction, since different distances between the laser
shots and the telescopes can be sampled, as opposed to the standard
analysis, in which the distance to the central laser facility is fixed
(see Ref.~\cite{PierreAuger:2021obf} for a preliminary demonstration
of this method).  A second opportunity stems from the public ATLID
aerosol products, which supply aerosol backscatter and extinction
coefficients with 300~m horizontal resolution and vertical resolution
of 100~m up to 20~km, and 500~m up to 40~km.

\section{Summary and Outlook}

In these proceedings we reported the detection of spaceborne lasers
with the Pierre Auger Observatory and presented a detailed analysis of
lidar shots from the ALADIN instrument aboard Aeolus and reported on
first observations of ATLID laser tracks from EarthCARE.

The successful ground-truthing of ATLID highlights the value of
astroparticle observatories for validating and calibrating space-based
lidar missions.  Even more intriguing, from the perspective
of ultra-high-energy cosmic-ray studies, is the reciprocal benefit:
the lasers aboard EarthCARE and future missions provide an independent
check on the aerosol profiles above each array.  Acting as ``standard
candles,'' they enable a global cross-calibration of cosmic-ray
observatories and help reduce current uncertainties in the full-sky
flux and anisotropy~\cite{PierreAuger:2023mvf,PierreAuger:2023mvk}. This method of cross-calibrate a world-wide observatory can also be applied to the Cherenkov Telscope Array Observatory (CTAO) \cite{CTAConsortium:2010umy} and, using future spaceborne lidars, to the Global Cosmic Ray Observatory (GCOS)~\cite{Ahlers:2025pqg}.

\footnotesize
\setlength{\bibsep}{3pt plus 0ex}
\begin{multicols}{2}
  \setlength\columnsep{3pt}
 \setstretch{0.9}
  \bibliographystyle{uhecr}
\bibliography{main}
\end{multicols}

\newpage

\section*{The Pierre Auger Collaboration}

{\footnotesize\setlength{\baselineskip}{10pt}
\noindent
\begin{wrapfigure}[11]{l}{0.12\linewidth}
\vspace{-4pt}
\includegraphics[width=0.98\linewidth]{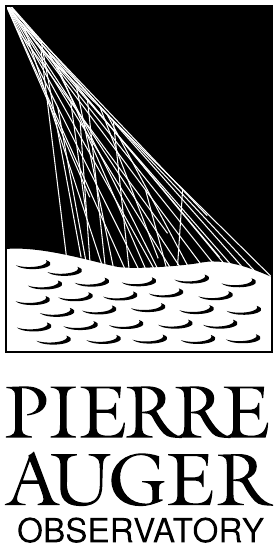}
\end{wrapfigure}
\begin{sloppypar}\noindent
A.~Abdul Halim$^{13}$,
P.~Abreu$^{70}$,
M.~Aglietta$^{53,51}$,
I.~Allekotte$^{1}$,
K.~Almeida Cheminant$^{78,77}$,
A.~Almela$^{7,12}$,
R.~Aloisio$^{44,45}$,
J.~Alvarez-Mu\~niz$^{76}$,
A.~Ambrosone$^{44}$,
J.~Ammerman Yebra$^{76}$
% ...

\end{sloppypar}
\begin{center}
\end{center}

\vspace{1ex}
\begin{description}[labelsep=0.2em,align=right,labelwidth=0.7em,labelindent=0em,leftmargin=2em,noitemsep,before={\renewcommand\makelabel[1]{##1 }}]
\item[$^{1}$] Centro At\'omico Bariloche and Instituto Balseiro (CNEA-UNCuyo-CONICET), San Carlos de Bariloche, Argentina
\item[$^{2}$] Departamento de F\'\i{}sica and Departamento de Ciencias de la Atm\'osfera y los Oc\'eanos, FCEyN, Universidad de Buenos Aires and CONICET, Buenos Aires, Argentina
\item[$^{3}$] IFLP, Universidad Nacional de La Plata and CONICET, La Plata, Argentina
\item[$^{4}$] Instituto de Astronom\'\i{}a y F\'\i{}sica del Espacio (IAFE, CONICET-UBA), Buenos Aires, Argentina
\item[$^{5}$] Instituto de F\'\i{}sica de Rosario (IFIR) -- CONICET/U.N.R.\ and Facultad de Ciencias Bioqu\'\i{}micas y Farmac\'euticas U.N.R., Rosario, Argentina
\item[$^{6}$] Instituto de Tecnolog\'\i{}as en Detecci\'on y Astropart\'\i{}culas (CNEA, CONICET, UNSAM), and Universidad Tecnol\'ogica Nacional -- Facultad Regional Mendoza (CONICET/CNEA), Mendoza, Argentina
\item[$^{7}$] Instituto de Tecnolog\'\i{}as en Detecci\'on y Astropart\'\i{}culas (CNEA, CONICET, UNSAM), Buenos Aires, Argentina
\item[$^{8}$] International Center of Advanced Studies and Instituto de Ciencias F\'\i{}sicas, ECyT-UNSAM and CONICET, Campus Miguelete -- San Mart\'\i{}n, Buenos Aires, Argentina
\item[$^{9}$] Laboratorio Atm\'osfera -- Departamento de Investigaciones en L\'aseres y sus Aplicaciones -- UNIDEF (CITEDEF-CONICET), Argentina
\item[$^{10}$] Observatorio Pierre Auger, Malarg\"ue, Argentina
\item[$^{11}$] Observatorio Pierre Auger and Comisi\'on Nacional de Energ\'\i{}a At\'omica, Malarg\"ue, Argentina
\item[$^{12}$] Universidad Tecnol\'ogica Nacional -- Facultad Regional Buenos Aires, Buenos Aires, Argentina
\item[$^{13}$] University of Adelaide, Adelaide, S.A., Australia
\item[$^{14}$] Universit\'e Libre de Bruxelles (ULB), Brussels, Belgium
\item[$^{15}$] Vrije Universiteit Brussels, Brussels, Belgium
\item[$^{16}$] Centro Brasileiro de Pesquisas Fisicas, Rio de Janeiro, RJ, Brazil
\item[$^{17}$] Centro Federal de Educa\c{c}\~ao Tecnol\'ogica Celso Suckow da Fonseca, Petropolis, Brazil
\item[$^{18}$] Instituto Federal de Educa\c{c}\~ao, Ci\^encia e Tecnologia do Rio de Janeiro (IFRJ), Brazil
\item[$^{19}$] Universidade de S\~ao Paulo, Escola de Engenharia de Lorena, Lorena, SP, Brazil
\item[$^{20}$] Universidade de S\~ao Paulo, Instituto de F\'\i{}sica de S\~ao Carlos, S\~ao Carlos, SP, Brazil
\item[$^{21}$] Universidade de S\~ao Paulo, Instituto de F\'\i{}sica, S\~ao Paulo, SP, Brazil
\item[$^{22}$] Universidade Estadual de Campinas (UNICAMP), IFGW, Campinas, SP, Brazil
\item[$^{23}$] Universidade Estadual de Feira de Santana, Feira de Santana, Brazil
\item[$^{24}$] Universidade Federal de Campina Grande, Centro de Ciencias e Tecnologia, Campina Grande, Brazil
\item[$^{25}$] Universidade Federal do ABC, Santo Andr\'e, SP, Brazil
\item[$^{26}$] Universidade Federal do Paran\'a, Setor Palotina, Palotina, Brazil
\item[$^{27}$] Universidade Federal do Rio de Janeiro, Instituto de F\'\i{}sica, Rio de Janeiro, RJ, Brazil
\item[$^{28}$] Universidad de Medell\'\i{}n, Medell\'\i{}n, Colombia
\item[$^{29}$] Universidad Industrial de Santander, Bucaramanga, Colombia
\item[$^{30}$] Charles University, Faculty of Mathematics and Physics, Institute of Particle and Nuclear Physics, Prague, Czech Republic
\item[$^{31}$] Institute of Physics of the Czech Academy of Sciences, Prague, Czech Republic
\item[$^{32}$] Palacky University, Olomouc, Czech Republic
\item[$^{33}$] CNRS/IN2P3, IJCLab, Universit\'e Paris-Saclay, Orsay, France
\item[$^{34}$] Laboratoire de Physique Nucl\'eaire et de Hautes Energies (LPNHE), Sorbonne Universit\'e, Universit\'e de Paris, CNRS-IN2P3, Paris, France
\item[$^{35}$] Univ.\ Grenoble Alpes, CNRS, Grenoble Institute of Engineering Univ.\ Grenoble Alpes, LPSC-IN2P3, 38000 Grenoble, France
\item[$^{36}$] Universit\'e Paris-Saclay, CNRS/IN2P3, IJCLab, Orsay, France
\item[$^{37}$] Bergische Universit\"at Wuppertal, Department of Physics, Wuppertal, Germany
\item[$^{38}$] Karlsruhe Institute of Technology (KIT), Institute for Experimental Particle Physics, Karlsruhe, Germany
\item[$^{39}$] Karlsruhe Institute of Technology (KIT), Institut f\"ur Prozessdatenverarbeitung und Elektronik, Karlsruhe, Germany
\item[$^{40}$] Karlsruhe Institute of Technology (KIT), Institute for Astroparticle Physics, Karlsruhe, Germany
\item[$^{41}$] RWTH Aachen University, III.\ Physikalisches Institut A, Aachen, Germany
\item[$^{42}$] Universit\"at Hamburg, II.\ Institut f\"ur Theoretische Physik, Hamburg, Germany
\item[$^{43}$] Universit\"at Siegen, Department Physik -- Experimentelle Teilchenphysik, Siegen, Germany
\item[$^{44}$] Gran Sasso Science Institute, L'Aquila, Italy
\item[$^{45}$] INFN Laboratori Nazionali del Gran Sasso, Assergi (L'Aquila), Italy
\item[$^{46}$] INFN, Sezione di Catania, Catania, Italy
\item[$^{47}$] INFN, Sezione di Lecce, Lecce, Italy
\item[$^{48}$] INFN, Sezione di Milano, Milano, Italy
\item[$^{49}$] INFN, Sezione di Napoli, Napoli, Italy
\item[$^{50}$] INFN, Sezione di Roma ``Tor Vergata'', Roma, Italy
\item[$^{51}$] INFN, Sezione di Torino, Torino, Italy
\item[$^{52}$] Istituto di Astrofisica Spaziale e Fisica Cosmica di Palermo (INAF), Palermo, Italy
\item[$^{53}$] Osservatorio Astrofisico di Torino (INAF), Torino, Italy
\item[$^{54}$] Politecnico di Milano, Dipartimento di Scienze e Tecnologie Aerospaziali , Milano, Italy
\item[$^{55}$] Universit\`a del Salento, Dipartimento di Matematica e Fisica ``E.\ De Giorgi'', Lecce, Italy
\item[$^{56}$] Universit\`a dell'Aquila, Dipartimento di Scienze Fisiche e Chimiche, L'Aquila, Italy
\item[$^{57}$] Universit\`a di Catania, Dipartimento di Fisica e Astronomia ``Ettore Majorana``, Catania, Italy
\item[$^{58}$] Universit\`a di Milano, Dipartimento di Fisica, Milano, Italy
\item[$^{59}$] Universit\`a di Napoli ``Federico II'', Dipartimento di Fisica ``Ettore Pancini'', Napoli, Italy
\item[$^{60}$] Universit\`a di Palermo, Dipartimento di Fisica e Chimica ''E.\ Segr\`e'', Palermo, Italy
\item[$^{61}$] Universit\`a di Roma ``Tor Vergata'', Dipartimento di Fisica, Roma, Italy
\item[$^{62}$] Universit\`a Torino, Dipartimento di Fisica, Torino, Italy
\item[$^{63}$] Benem\'erita Universidad Aut\'onoma de Puebla, Puebla, M\'exico
\item[$^{64}$] Unidad Profesional Interdisciplinaria en Ingenier\'\i{}a y Tecnolog\'\i{}as Avanzadas del Instituto Polit\'ecnico Nacional (UPIITA-IPN), M\'exico, D.F., M\'exico
\item[$^{65}$] Universidad Aut\'onoma de Chiapas, Tuxtla Guti\'errez, Chiapas, M\'exico
\item[$^{66}$] Universidad Michoacana de San Nicol\'as de Hidalgo, Morelia, Michoac\'an, M\'exico
\item[$^{67}$] Universidad Nacional Aut\'onoma de M\'exico, M\'exico, D.F., M\'exico
\item[$^{68}$] Institute of Nuclear Physics PAN, Krakow, Poland
\item[$^{69}$] University of \L{}\'od\'z, Faculty of High-Energy Astrophysics,\L{}\'od\'z, Poland
\item[$^{70}$] Laborat\'orio de Instrumenta\c{c}\~ao e F\'\i{}sica Experimental de Part\'\i{}culas -- LIP and Instituto Superior T\'ecnico -- IST, Universidade de Lisboa -- UL, Lisboa, Portugal
\item[$^{71}$] ``Horia Hulubei'' National Institute for Physics and Nuclear Engineering, Bucharest-Magurele, Romania
\item[$^{72}$] Institute of Space Science, Bucharest-Magurele, Romania
\item[$^{73}$] Center for Astrophysics and Cosmology (CAC), University of Nova Gorica, Nova Gorica, Slovenia
\item[$^{74}$] Experimental Particle Physics Department, J.\ Stefan Institute, Ljubljana, Slovenia
\item[$^{75}$] Universidad de Granada and C.A.F.P.E., Granada, Spain
\item[$^{76}$] Instituto Galego de F\'\i{}sica de Altas Enerx\'\i{}as (IGFAE), Universidade de Santiago de Compostela, Santiago de Compostela, Spain
\item[$^{77}$] IMAPP, Radboud University Nijmegen, Nijmegen, The Netherlands
\item[$^{78}$] Nationaal Instituut voor Kernfysica en Hoge Energie Fysica (NIKHEF), Science Park, Amsterdam, The Netherlands
\item[$^{79}$] Stichting Astronomisch Onderzoek in Nederland (ASTRON), Dwingeloo, The Netherlands
\item[$^{80}$] Universiteit van Amsterdam, Faculty of Science, Amsterdam, The Netherlands
\item[$^{81}$] Case Western Reserve University, Cleveland, OH, USA
\item[$^{82}$] Colorado School of Mines, Golden, CO, USA
\item[$^{83}$] Department of Physics and Astronomy, Lehman College, City University of New York, Bronx, NY, USA
\item[$^{84}$] Michigan Technological University, Houghton, MI, USA
\item[$^{85}$] New York University, New York, NY, USA
\item[$^{86}$] University of Chicago, Enrico Fermi Institute, Chicago, IL, USA
\item[$^{87}$] University of Delaware, Department of Physics and Astronomy, Bartol Research Institute, Newark, DE, USA
\item[] -----
\item[$^{a}$] Max-Planck-Institut f\"ur Radioastronomie, Bonn, Germany
\item[$^{b}$] also at Kapteyn Institute, University of Groningen, Groningen, The Netherlands
\item[$^{c}$] School of Physics and Astronomy, University of Leeds, Leeds, United Kingdom
\item[$^{d}$] Fermi National Accelerator Laboratory, Fermilab, Batavia, IL, USA
\item[$^{e}$] Pennsylvania State University, University Park, PA, USA
\item[$^{f}$] Colorado State University, Fort Collins, CO, USA
\item[$^{g}$] Louisiana State University, Baton Rouge, LA, USA
\item[$^{h}$] now at Graduate School of Science, Osaka Metropolitan University, Osaka, Japan
\item[$^{i}$] Institut universitaire de France (IUF), France
\item[$^{j}$] now at Technische Universit\"at Dortmund and Ruhr-Universit\"at Bochum, Dortmund and Bochum, Germany
\end{description}

\section*{Acknowledgments}

\begin{sloppypar}
The successful installation, commissioning, and operation of the Pierre
Auger Observatory would not have been possible without the strong
commitment and effort from the technical and administrative staff in
Malarg\"ue. We are very grateful to the following agencies and
organizations for financial support:
\end{sloppypar}

\begin{sloppypar}
Argentina -- Comisi\'on Nacional de Energ\'\i{}a At\'omica; Agencia Nacional de
Promoci\'on Cient\'\i{}fica y Tecnol\'ogica (ANPCyT); Consejo Nacional de
Investigaciones Cient\'\i{}ficas y T\'ecnicas (CONICET); Gobierno de la
Provincia de Mendoza; Municipalidad de Malarg\"ue; NDM Holdings and Valle
Las Le\~nas; in gratitude for their continuing cooperation over land
access; Australia -- the Australian Research Council; Belgium -- Fonds
de la Recherche Scientifique (FNRS); Research Foundation Flanders (FWO),
Marie Curie Action of the European Union Grant No.~101107047; Brazil --
Conselho Nacional de Desenvolvimento Cient\'\i{}fico e Tecnol\'ogico (CNPq);
Financiadora de Estudos e Projetos (FINEP); Funda\c{c}\~ao de Amparo \`a
Pesquisa do Estado de Rio de Janeiro (FAPERJ); S\~ao Paulo Research
Foundation (FAPESP) Grants No.~2019/10151-2, No.~2010/07359-6 and
No.~1999/05404-3; Minist\'erio da Ci\^encia, Tecnologia, Inova\c{c}\~oes e
Comunica\c{c}\~oes (MCTIC); Czech Republic -- GACR 24-13049S, CAS LQ100102401,
MEYS LM2023032, CZ.02.1.01/0.0/0.0/16{\textunderscore}013/0001402,
CZ.02.1.01/0.0/0.0/18{\textunderscore}046/0016010 and
CZ.02.1.01/0.0/0.0/17{\textunderscore}049/0008422 and CZ.02.01.01/00/22{\textunderscore}008/0004632;
France -- Centre de Calcul IN2P3/CNRS; Centre National de la Recherche
Scientifique (CNRS); Conseil R\'egional Ile-de-France; D\'epartement
Physique Nucl\'eaire et Corpusculaire (PNC-IN2P3/CNRS); D\'epartement
Sciences de l'Univers (SDU-INSU/CNRS); Institut Lagrange de Paris (ILP)
Grant No.~LABEX ANR-10-LABX-63 within the Investissements d'Avenir
Programme Grant No.~ANR-11-IDEX-0004-02; Germany -- Bundesministerium
f\"ur Bildung und Forschung (BMBF); Deutsche Forschungsgemeinschaft (DFG);
Finanzministerium Baden-W\"urttemberg; Helmholtz Alliance for
Astroparticle Physics (HAP); Helmholtz-Gemeinschaft Deutscher
Forschungszentren (HGF); Ministerium f\"ur Kultur und Wissenschaft des
Landes Nordrhein-Westfalen; Ministerium f\"ur Wissenschaft, Forschung und
Kunst des Landes Baden-W\"urttemberg; Italy -- Istituto Nazionale di
Fisica Nucleare (INFN); Istituto Nazionale di Astrofisica (INAF);
Ministero dell'Universit\`a e della Ricerca (MUR); CETEMPS Center of
Excellence; Ministero degli Affari Esteri (MAE), ICSC Centro Nazionale
di Ricerca in High Performance Computing, Big Data and Quantum
Computing, funded by European Union NextGenerationEU, reference code
CN{\textunderscore}00000013; M\'exico -- Consejo Nacional de Ciencia y Tecnolog\'\i{}a
(CONACYT) No.~167733; Universidad Nacional Aut\'onoma de M\'exico (UNAM);
PAPIIT DGAPA-UNAM; The Netherlands -- Ministry of Education, Culture and
Science; Netherlands Organisation for Scientific Research (NWO); Dutch
national e-infrastructure with the support of SURF Cooperative; Poland
-- Ministry of Education and Science, grants No.~DIR/WK/2018/11 and
2022/WK/12; National Science Centre, grants No.~2016/22/M/ST9/00198,
2016/23/B/ST9/01635, 2020/39/B/ST9/01398, and 2022/45/B/ST9/02163;
Portugal -- Portuguese national funds and FEDER funds within Programa
Operacional Factores de Competitividade through Funda\c{c}\~ao para a Ci\^encia
e a Tecnologia (COMPETE); Romania -- Ministry of Research, Innovation
and Digitization, CNCS-UEFISCDI, contract no.~30N/2023 under Romanian
National Core Program LAPLAS VII, grant no.~PN 23 21 01 02 and project
number PN-III-P1-1.1-TE-2021-0924/TE57/2022, within PNCDI III; Slovenia
-- Slovenian Research Agency, grants P1-0031, P1-0385, I0-0033, N1-0111;
Spain -- Ministerio de Ciencia e Innovaci\'on/Agencia Estatal de
Investigaci\'on (PID2019-105544GB-I00, PID2022-140510NB-I00 and
RYC2019-027017-I), Xunta de Galicia (CIGUS Network of Research Centers,
Consolidaci\'on 2021 GRC GI-2033, ED431C-2021/22 and ED431F-2022/15),
Junta de Andaluc\'\i{}a (SOMM17/6104/UGR and P18-FR-4314), and the European
Union (Marie Sklodowska-Curie 101065027 and ERDF); USA -- Department of
Energy, Contracts No.~DE-AC02-07CH11359, No.~DE-FR02-04ER41300,
No.~DE-FG02-99ER41107 and No.~DE-SC0011689; National Science Foundation,
Grant No.~0450696, and NSF-2013199; The Grainger Foundation; Marie
Curie-IRSES/EPLANET; European Particle Physics Latin American Network;
and UNESCO.
\end{sloppypar}

}

\end{document}